\documentclass{aa}

\usepackage{txfonts}
\usepackage{graphicx}

\begin{document}

\title{Influence of UV radiation from a massive YSO on the chemistry of its envelope}
\author{P. St\"auber\inst{1} \and S.D. Doty\inst{3} \and E.F. van Dishoeck\inst{2} \and J.K. J{\o}rgensen\inst{2} \and A.O. Benz\inst{1}}
\offprints{P. St\"auber, \email{pascalst@astro.phys.ethz.ch}}
\institute{Institute of Astronomy, ETH-Zentrum, CH-8092 Zurich, Switzerland \and Sterrewacht Leiden, PO Box 9513, 2300 RA Leiden, The Netherlands \and Department of Physics and Astronomy, Denison University, Granville, OH 43023, USA}
\date{Received / Accepted }

\abstract{We have studied the influence of far ultraviolet (UV)
radiation ($6$$<$ $h\nu$ $<$ $13.6$\,eV) from a massive young stellar
object (YSO) on the chemistry of its own envelope by extending the
models of Doty et al. (\cite{Doty}) to include a central source of UV 
radiation. The models are applied to the massive star-forming region
AFGL $2591$ for different inner UV field strengths. Depth-dependent
abundance profiles for several molecules are presented and
discussed. We predict enhanced column densities for more than $30$
species, especially radicals and ions.  Comparison between
observations and models is improved with a moderate UV field incident
on the inner envelope, corresponding to an enhancement factor $G_{0}
\approx 10$--$100$ at $200$\,AU from the star with an optical depth
$\tau \approx 15$--$17$. The chemical networks of various
species are explored. Subtle differences are found compared with
traditional models of Photon Dominated Regions (PDRs) because of the
higher temperatures and higher gas-phase H$_2$O abundance caused by
evaporation of ices in the inner region. In particular, the
\element{CN}/\element{HCN} ratio is not a sensitive tracer of the
inner UV field, in contrast with the situation for normal PDRs: for
low UV fields, the extra CN reacts with H$_2$ in the inner dense and
warm region and produces more HCN. It is found that the \element{CH^+}
abundance is strongly enhanced and grows steadily with increasing UV
field. In addition, the ratio \element{CH^+}/\element{CH} is increased
by a factor of $10^3$--$10^5$ depending on the inner UV
flux. High-$J$ lines of molecules like CN and HCN are most sensitive
to the inner dense region where UV radiation plays a role. Thus, even
though the total column density affected by UV photons is small,
comparison of high-$J$ and low-$J$ lines can selectively trace and
distinguish the inner UV field from the outer one. In addition, future
Herschel-HIFI observations of hydrides can sensitively probe the inner
UV field.  }

\maketitle

\titlerunning{Influence of UV radiation on massive YSO envelopes}
\authorrunning{P. St\"auber et al.}

\section{Introduction}

A newly formed massive star in the earliest stages of evolution is
surrounded by a dense envelope and is deeply embedded in its natal
molecular cloud. Although highly obscured, these objects show signs of
different stellar activity. Radio continuum observations reveal
compact \ion{H}{ii} regions close to the young star (e.g., Churchwell
\cite{Churchwell}) and molecular outflows have been observed toward
numerous massive star-forming regions (Shepherd \& Churchwell
\cite{Shepherd}; Beuther et al. \cite{Beuther}). Outflows and high
energy photons from the YSO can strongly affect both the physical and
chemical structure of the envelope. Observational studies of massive
YSO environments reveal rich molecular spectra with a large variety of
species (see e.g., Blake et al. \cite{Blake87}, \cite{Blake96}; 
Turner \cite{Turner}; Helmich \& van Dishoeck \cite{Helmich}; Gibb et
al. \cite{Gibb00}). This molecular complexity is thought to be
associated with a specific stage in the evolution of a massive young
star ---the so-called `hot core' phase--- in which selected molecules
have just evaporated off the grains and drive a rich high temperature
chemistry. However, there are also massive YSOs which do not show this
molecular complexity, but where energetic phenomena can still cause
chemical changes.  Indeed, chemistry may be a unique diagnostic of the
relative importance of different physical processes that occur within
the YSO envelope, and a good understanding of chemistry is therefore
crucial to comprehend the birth and evolution of massive stars.

Much effort has been put into the quest of understanding the physical
and chemical structure of massive YSO envelopes in recent years. Van
der Tak et al. (\cite{Tak1999}, \cite{Tak2000}), Mueller et
al. (\cite{Mue02}) and Beuther et al. (\cite{Beu02b}) investigated the
physical structure of the envelopes around several massive young stars
on $100$--$10^5$\,AU scales, using lines of the CS molecule and
continuum data.  Adopting the resulting density profiles as input,
Doty et al. (\cite{Doty}) calculated detailed gas temperature and
chemical models to test these physical structures and, in particular,
to study the chemical evolution of the high-mass star forming region
AFGL $2591$. Self-consistent models of the gas and dust thermal
balance, chemistry, and radiative transfer within YSO envelopes have
also been performed by Doty \& Neufeld (\cite{DotyN}) for
high-mass objects and Ceccarelli et al. (\cite{Ceccarellib}) for
low-mass sources. Most recently, Rodgers \& Charnley (\cite{Rodgers})
have modeled the chemistry of gravitationally collapsing low-mass YSO
envelopes considering spatial density variations and gradual heating
of the gas as well as temporal chemical evolution.  

All of these models do not consider a central source of UV radiation,
and include (at most) an external radiation field comparable to the
average interstellar radiation field. The only exception is the work
by Ceccarelli et al. (\cite{Ceccarellia}), who modeled the chemistry
around a low-mass YSO with central UV radiation and calculated radial
abundances between $\approx 1000$--$10^{5}$\,AU. Their assumed
density profile was flat rather than a power-law, whereby UV
photons can penetrate further. The results from such models resemble
those from standard models of photon-dominated regions (PDRs), in
which the UV radiation from a hot star impacts a nearby molecular
cloud. In these regions the physical and chemical structure is
determined by the incident FUV field ($6$\,eV $<$ $h\nu$ $<$
$13.6$\,eV; see e.g., Sternberg \& Dalgarno \cite{Sternberg},
Hollenbach \& Tielens \cite{Hollenbach}). PDRs are characterized by
bright atomic fine-structure lines, strong high$-J$ CO emission, PAHs
and specific molecules like CN, C$_2$H or CO$^+$ (e.g., Jansen et
al. \cite{Jansen}, St\"orzer et al. \cite{Storzer}).  Thus, while
there is no doubt that UV photons affect their surroundings in the
later stages of massive star evolution when the envelope is being
dispersed, their importance in the early stages is still unclear.

The question whether or not UV radiation plays a role in the chemistry
of YSO envelopes has been a hotly debated subject for some time.
Because of the limited penetration depth of UV photons (typically a
few magnitudes of visual extinction $A_V$), the bulk of the envelope
with $A_V>100$ mag must be unaffected. However, the presence of
certain solid-state features, in particular the XCN or OCN$^-$ band at
$4.62$\,$\mu$m, is often cited as an indicator of `energetic
processing' (e.g., Pendleton et al. \cite{Pendleton}). Whether this
indeed implies the presence of UV radiation or simply requires higher
temperatures is still uncertain. Indeed, recent laboratory experiments
by van Broekhuizen et al. (\cite{Broekhuizen}) indicate that similar
production efficiencies of OCN$^-$ can be obtained with thermal
heating of HNCO-containing mixtures, reducing its diagnostic
value. PAH emission is not detected in the earliest embedded stages,
but this could be due either to freezing out of the PAHs onto grains
or to the high extinction at infrared wavelengths. The aim of this
paper is therefore to find other diagnostics of the inner UV field, in
particular from molecular tracers whose (sub-)millimeter emission can
penetrate the entire envelope.

We study here the chemistry in YSO envelopes under the influence of UV
radiation from the central source as well as from an outside
interstellar FUV radiation field. For this purpose, we have extended
the time- and position-dependent detailed chemical model of Doty et
al. (\cite{Doty}) to allow the impact of a central UV field on the
inner envelope. In addition, we have used the resulting abundance
profiles to compute emission lines of selected species using the Monte
Carlo code of Hogerheijde \& van der Tak (\cite{Hogerheijde}), for
direct comparison with observations.  Although our applications are
focused on high-mass YSOs, the results should be equally applicable
to the envelopes around low-mass sources, albeit that the origin of
the UV radiation may be slightly different.

This paper is organized as follows. In Sect.~\ref{pm}, the physical
and chemical model is described. The model is then applied to the
envelope around AFGL $2591$ and the results are discussed. We examine
the highly abundant molecules \element{CO}, \element{H_2O} and
\element{CO_2} in Sect.~\ref{mjm}. In Sect.~\ref{ppdr} we discuss the
`PDR-related' elements and ions \element{C}, \element{O} and
\element{C^+}. The reactive ions \element{CO^+}, recently observed
toward the high-mass star forming region W$3$ IRS$5$ by St\"auber et
al. (\cite{Stp}), \element{SO^+}, \element{HCO^+} and \element{HOC^+},
which were also studied by Fuente et al. (\cite{Fuente}) in PDRs, are
examined in Sect.~\ref{prion}. Hydrides and ions like \element{OH},
\element{OH^+}, \element{CH} and \element{CH^+} are discussed in
Sect.~\ref{phifi}. These are important species for future space-borne
observatories like Herschel where they will be observable with the
high resolution spectrometer HIFI (de Graauw \& Helmich
\cite{HIFI}). In Sect.~\ref{pcn}, \element{CN}-bearing molecules are
studied since they are thought to be sensitive to UV radiation and
readily observable with existing instrumentation. Line emission of
\element{CN} and \element{HC^{15}N} is calculated and the results 
are presented and discussed in Sect.~\ref{pline}. 
We conclude by summarizing our results in Sect.~\ref{pcon}.

\section{Model}
\label{pm}

\begin{figure}
\centering
\resizebox{\hsize}{!}{\includegraphics{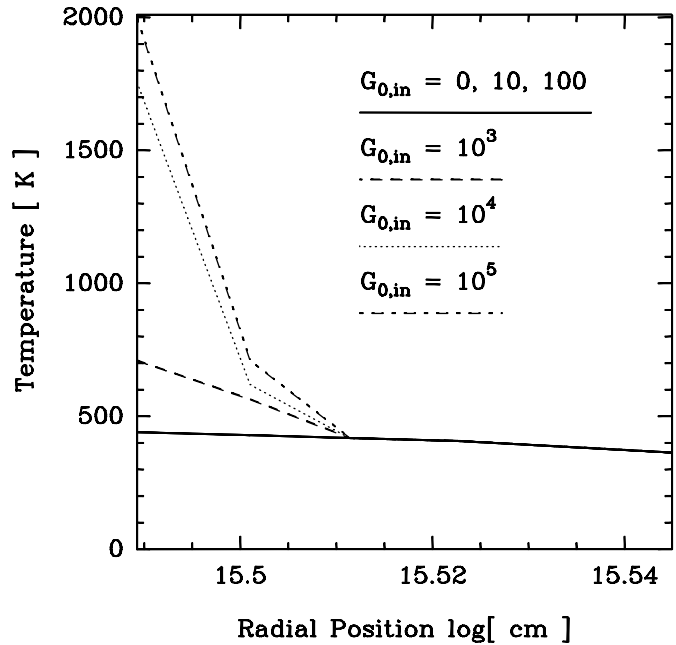}}
\caption{Gas temperature distribution in the inner $200$--$250$\,AU for different UV fluxes 
$G_{0,\rm{in}}$. The dust temperature $T_{dust}$ equals the gas temperature for low
UV fields ($G_{0} \le 100$).}
\label{ftemp}
\end{figure}

\begin{figure}
\centering
\resizebox{\hsize}{!}{\includegraphics{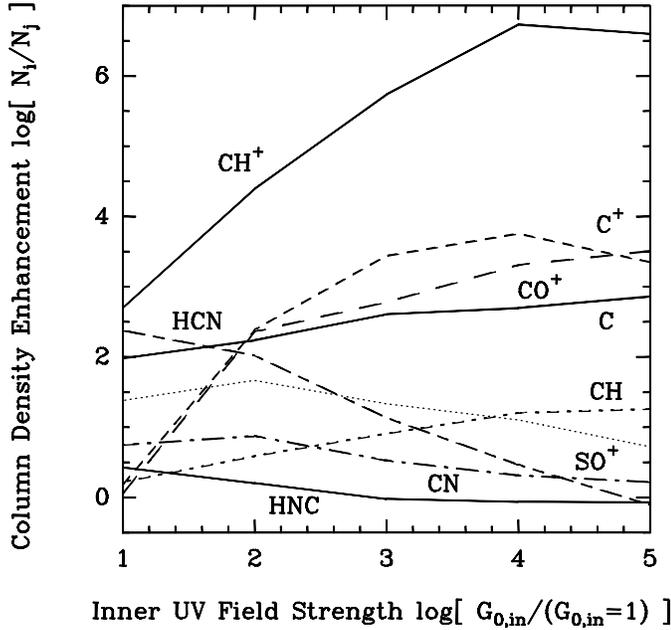}}
\caption{Column densities of various species as functions of $G_{0,\rm{in}}$, normalized to the value for $G_{0,\rm{in}} = 0$.}
\label{fdens}
\end{figure}

In order to calculate the abundances of each species at a certain time
and distance from the central object, a physical and thermal model of
the envelope is needed and initial conditions, i.e., initial
abundances, have to be assumed. We adopt the values provided by Doty
et al. (\cite{Doty}) and refer to that paper and references therein
for more detailed information. However, a brief outline is given in
this section.

\subsection{Physical and thermal model}

Van der Tak et al. (\cite{Tak1999}, \cite{Tak2000}) fitted their
observations of AFGL $2591$ with a spherical power-law density
distribution $n = n_{0}(r_{0}/r)^{\alpha}$ ($n_{0} = 5.8 \times10^4$\,cm$^{-3}$,
$r_{0} = 2.7 \times 10^4$\,AU, $\alpha = 1.0$) by
modeling CS emission lines over a large range of critical densities
and fitting the continuum emission. Our model covers a region that extends 
from $(r_{\rm{in}},n_{\rm in}) \approx (200$\,AU$,5.9 \times 10^{6}$\,cm$^{-3})$ to 
$(r_{\rm{out}},n_{\rm out})\approx (29\,000$\,AU$,4.2 \times 10^{4}$\,cm$^{-3})$. 
The dust temperature profile was determined from a self-consistent solution of the 
continuum radiative transfer problem. As shown by Doty \& Neufeld (\cite{DotyN}) and Doty
et al. (\cite{Doty}), the assumption that the gas and dust
temperatures are equal, $T_{\rm{gas}} \approx T_{\rm{dust}}$, holds
throughout most of the envelope  because of the high density. However, 
at the inner edge, photoelectric heating and other
heating processes proportional to the radiation field (e.g.,
collisional de-excitation of vibrationally excited H$_2$ pumped by FUV
radiation) can become important (Tielens \& Hollenbach \cite{Tielens}, Sternberg
\& Dalgarno \cite{Sternberg89}). Comparison of these heating rates with the cooling
rate due to gas-dust collisions (Hollenbach \& McKee \cite{HoK}) shows that
the difference in gas and dust temperatures is only a few K for low
FUV fields ($G_{\rm 0,in}\leq 100$; see Sect.~\ref{pchm} and \ref{puvf} for
definition). For higher FUV fields, the gas temperature becomes
significantly higher, with values up to a few thousand K at the edge.

Figure~\ref{ftemp} illustrates the gas temperature in the inner region computed
taking the above-mentioned processes into account for values of
$G_{\rm 0,in}$ up to $10^5$. The gas temperature drops quickly within
$A_V\leq 1$ and equals the dust temperature further in the cloud.  The
dust temperature ---in particular the freeze-out radius ($T_{\rm
dust}\approx 100$\,K)--- is not affected beyond $A_V\approx 1$, since
the FUV photons are quickly re-distributed in frequency and the total
luminosity of the system is unchanged (Doty \& Leung \cite{DotyL}). 
The physical structure is kept constant with time. Following the arguments 
summarized in van der Tak et al. (\cite{Tak1999}), we assume a distance to 
AFGL $2591$ of $1$\,kpc. 

\subsection{Chemical model}
\label{pchm}

The chemical model is based upon the UMIST gas-phase chemical reaction
network (Millar et al. \cite{Millar}) and basically solves the
equations for molecular evolution given by
\begin{equation}
\frac{dn(i)}{dt} = \sum_{j} k_{ij} n(j) + \sum_{j,k} l_{ijk} n(j) n(k)
\label{meq}
\end{equation}
where $n(i)$ is the number density of species $i$; $k_{ij}$ and
$l_{ijk}$ are the rate coefficients for production and destruction of
the given species. The main inputs of the model are the temperature
and density distribution, the cosmic ray ionization rate $\zeta
\approx 5.6 \times 10^{-17}$\,s$^{-1}$ for AFGL $2591$, and the strength of
the FUV field $G_{0}$ with respect to the average interstellar flux at
$6 < h\nu < 13.6$\,eV of $1.6 \times 10^{-3}$\,erg\,cm$^{-2}$\,s$^{-1}$.  
The model then calculates the
time- and depth-dependent chemistry for $395$ species and $3865$
reactions. The chemical abundances are calculated over a range of $34$
radial grid points, providing a time- and space-dependent chemical evolution. The
approximate effects of freeze-out onto dust grains are included by
initially depleting certain species below $100$\,K. The effects of
changing the cosmic-ray ionization rate have been investigated by Doty
et al. (\cite{Doty}) and are not further discussed in this paper.

\subsection{UV radiation field}
\label{puvf}

In our spherical model, the envelope can be irradiated by UV photons
from the inside by a central source and from the outside by the
interstellar radiation field (ISRF). The UV flux is characterized in
units of $G_{0}$ (see Sect.~\ref{pchm}). We therefore denote the enhancement 
factor due to the inner source $G_{0,\rm{in}}$ and that connected to the ISRF
$G_{0,\rm{out}}$, assuming the same spectral energy distribution. The
rates for the photodissociation and photoionization
reactions can then be approximated by (see also van Dishoeck
\cite{Ewine})
\begin{equation}
k = G_{0,\rm{in}} C e^{-\gamma \tau_{\rm{in}}} + G_{0,\rm{out}} C e^{-\gamma \tau_{\rm{out}}} \quad \bigl[\rm{s}^{-1}\bigr]
\label{keq}
\end{equation}
where $C$ and $\gamma$ are fit parameters and $\tau_{\rm{in}}$ and
$\tau_{\rm{out}}$ are the attenuation factors, approximated by
$\tau \approx A_{V}$, the total visual extinction $A_{V}$.
The values of $A_V$ at each depth are calculated from the total
hydrogen column using the conversion factor $N(\element{H_{tot}}) =
2N(\element{H_{2}}) + N(\ion{H}{i}) \approx 1.86 \times 10^{21}
A_{V}$\,cm$^{-2}$, constrained from observations of translucent clouds
by Rachford et al. (\cite{Rachford}). For simplicity we start our
model with an attenuation of $\tau_{\rm{in}} = 0$ at $200$\, AU and
vary the inner UV flux from $G_{0,\rm{in}} = 0$ to $G_{0,\rm{in}} =
10^{5}$.  We have also carried out different models by varying
$G_{0,\rm{out}}$ from $0$ up to $100$ but the changes of the total
radial column densities due to different ISRF fluxes are very
small. The standard value for the ISRF flux in our models is therefore
$G_{0,\rm{out}} = 1$.

To obtain a rough indication of the inner UV field, we assume a
luminosity of $2 \times 10^{4}$\,L$_{\sun}$ and an effective
temperature of $3 \times 10^{4}$\,K for the central source as
suggested by van der Tak et al. (\cite{Tak1999}). A large fraction of
this luminosity may be due to accretion (Osorio et al. \cite{Osorio}).
This luminosity and stellar temperature would yield $G_{0,\rm{in}}
\approx 2 \times 10^{8}$ at $200$\,AU in the absence of any
absorption. Hence an estimated UV field strength of $G_0 \approx 10$
corresponds to a dust optical depth $\tau \approx 17$ at this
distance. A first question is whether such a relatively low
value for $\tau$ corresponds to a mass that is high enough to
provide a typical mass accretion rate for high mass YSOs of 
$\approx 10^{-4}$--$10^{-2}$\,M$_{\sun}$\,yr$^{-1}$ (e.g., Behrend \& Maeder \cite{Behrend}, McKee \& Tan \cite{McT}).
Using the relation between the total visual extinction and the hydrogen 
column density given earlier in this Sect., $A_{\rm{V}} \approx 17$\,mag corresponds to a column
density of $N(\rm{H_2}) \approx 2 \times 10^{22}$\,cm$^{-2}$ or a
density of $n \approx 10^{7}$\,cm$^{-3}$ at $200$\,AU. Assuming spherical free fall
inflow and a central mass of $M_{\ast} = 10$\,M$_{\sun}$, this density would be consistent
with an accretion rate of (e.g., Shu et al. \cite{Shu}) $\dot{M}_{\rm{acc}} = 4 \pi \rho r^{3/2}(2GM_{\ast})^{1/2} \approx 10^{-4}$\,M$_{\sun}$\,yr$^{-1}$.
It is expected that these high accretion rates `choke off' an incipient \ion{H}{ii}
region (Walmsley \cite{Walmsley}) and that the hydrogen-ionizing
photons ($E>13.6$\,eV) cannot travel far from the protostar before
being absorbed by infalling matter. The result would be a small
Str\"omgren sphere around the central protostar, with the non-ionizing
photons allowed to escape to larger distances. Both the luminosity and effective 
temperature of the radiation are actually expected to evolve during the protostellar 
stage (e.g., McKee \& Tan \cite{McT}), but such detailed modeling is beyond the scope 
of this study.   Note also that the geometry of this dust and inflowing material 
is not known and is therefore not included in our envelope model. If our model were 
extended inward to account for this extra material, only the first few radial positions
in our grid up to $A_V\approx 1$ would change.

Our models are spherically symmetric and do not include geometrical
effects like outflow cones or clumps in the central regions that could
allow UV photons to travel further into the envelope. The effects of
UV radiation escaping through outflow cones and scattering back on the
surrounding envelope have been studied by Spaans et al. (\cite{Spaans}) 
for the case of low-mass YSOs.  Finally, it should be mentioned that the 
photoreaction rates (\ref{keq}) are fitted for plane parallel models but 
are applied to spherical models. In spherical symmetry, the FUV photons can penetrate
further on average, but the gas temperature is somewhat lower assuming
the same density structure (St\"orzer et al. \cite{Storzer}). The authors conclude that
spherical geometry only becomes important for clouds with $n\approx 10^5$\,
cm $^{-3}$ exposed to FUV fields with intensities $G_0 \approx 10^3$ and a H$_2$ 
column density of less than $\approx 10^{22}$\,cm$^{-2}$. We therefore
neglect these effects.

\section{Chemistry results}
\label{result}

\begin{table*}
\centering
\caption[]{Predicted total radial column densities $N_{\rm{radial}}$ for different UV fluxes $G_{0,\rm{in}}$.}
\label{tcd}
\begin{tabular}{lcrcrcrcr} \hline\hline
Species & N [cm$^{-2}$] & Ratio & N [cm$^{-2}$] & Ratio & N [cm$^{-2}$] & Ratio & N [cm$^{-2}$] &  Ratio \\ 
        & $G_{0,\rm{in}} = 10$ & & $G_{0,\rm{in}} = 100$ & & $G_{0,\rm{in}} = 10^{3}$ & & $G_{0,\rm{in}} = 10^{5}$ & \\ \hline
          H & 4.5(19) &    5.9 & 1.3(20) &   17.6 & 2.3(20) & 30.1 & 1.4(21) &  187.1 \\
          C & 1.0(17) &   95.2 & 1.9(17) &  172.7 & 4.4(17) &  400.8 & 8.0(17) &  727.2 \\
          N & 7.4(16) &    2.1 & 1.4(17) &    3.9 & 2.8(17) & 8.0 & 4.9(17) &   14.0 \\
          O & 5.7(18) &    1.0 & 6.1(18) &    1.1 & 7.0(18) & 1.2 & 8.8(18) &    1.6 \\
         CH & 1.6(13) &    1.7 & 3.7(13) &    3.9 & 7.7(13) & 7.8 & 1.7(14) &   18.2 \\
     CH$_2$ & 1.2(14) &   59.3 & 2.0(13) &    9.6 & 1.7(13) & 8.3 & 6.6(13) &   32.1 \\
     CH$_3$ & 1.6(14) &    1.1 & 1.6(14) &    1.1 & 1.4(14) & 1.0 & 1.4(14) &    1.0 \\
         OH & 1.2(16) &    1.3 & 1.6(16) &    1.7 & 2.0(16) & 2.1 & 1.2(16) &    1.3 \\
      C$_2$ & 1.2(14) &    4.8 & 1.0(14) &    3.9 & 1.5(14) & 5.7 & 1.7(14) &    6.8 \\
        HCN & 1.4(17) &  238.6 & 6.3(16) &  105.5 & 8.2(15) & 13.7 & 4.7(14) &    0.8 \\
        HCO & 1.8(12) &    1.0 & 1.7(12) &    0.9 & 1.5(12) & 0.9 & 1.3(12) &    0.7 \\
        OCS & 4.0(14) &    1.5 & 1.1(14) &    0.4 & 7.1(13) & 0.3 & 4.9(13) &    0.2 \\
         CN & 3.1(14) &    5.6 & 4.2(14) &    7.5 & 1.8(14) & 3.3 & 9.2(13) &    1.7 \\
         CS & 1.0(15) &    2.4 & 7.9(14) &    1.9 & 4.2(14) & 1.0 & 3.7(14) &    0.9 \\
        HCS & 3.3(14) &   34.1 & 1.1(13) &    1.1 & 1.4(11) &  1(-2) & 1.5(11) &  2(-2) \\
     C$_2$H & 1.5(15) &  221.4 & 1.4(14) &   20.5 & 8.5(13) & 12.7 & 7.3(13) &   10.8 \\
      C$_3$ & 2.9(13) &    1.6 & 3.0(13) &    1.7 & 2.9(13) & 1.7 & 2.2(13) &    1.2 \\
     C$_3$N & 2.3(13) &    1.4 & 1.8(13) &    1.1 & 1.7(13) & 1.0 & 1.6(13) &    1.0 \\
       C$5$ & 7.0(12) &    1.3 & 7.3(12) &    1.4 & 6.8(12) & 1.3 & 5.9(12) &    1.1 \\
      C$5$N & 6.3(11) &    2.0 & 3.2(11) &    1.1 & 2.2(11) & 0.7 & 1.9(11) &    0.6 \\
        HNC & 2.0(15) &    2.7 & 1.2(15) &    1.6 & 7.3(14) & 1.0 & 6.5(14) &    0.9 \\
    HC$_3$N & 4.6(14) &    1.6 & 3.0(14) &    1.0 & 2.9(14) & 1.0 & 2.9(14) &    1.0 \\
     C$_2$S & 6.6(14) &   14.0 & 1.5(14) &    3.2 & 9.6(13) & 2.1 & 1.2(14) &    2.6 \\
     C$_3$S & 2.6(13) &    4.2 & 1.5(13) &    2.4 & 1.1(13) & 1.8 & 6.6(12) &    1.1 \\
   CH$_2$CN & 1.3(16) & 1074.3 & 6.8(15) &  566.0 & 6.2(14) & 51.9 & 7.4(12) &    0.6 \\ 
   CH$_3$CN & 8.9(13) &   28.7 & 6.1(13) &   19.9 & 1.2(13) & 3.8 & 2.7(12) &    0.9 \\
    H$_2$CS & 5.3(14) &    0.4 & 3.9(14) &    0.3 & 3.0(14) & 0.2 & 2.2(14) &    0.1 \\
      H$^+$ & 4.0(13) &    1.0 & 7.1(13) &    1.8 & 3.8(15) & 93.4 & 2.4(16) &  612.7 \\
      C$^+$ & 3.6(14) &    1.2 & 7.2(16) &  229.1 & 1.9(17) &  634.7 & 1.0(18) & 3220.0 \\
      O$^+$ & 3.1(08) &    1.3 & 7.9(09) &   31.4 & 6.0(11) & 2389.0 & 6.3(12) & 2.5(4) \\
      S$^+$ & 3.5(14) &  164.9 & 1.2(15) &  554.6 & 3.3(15) & 1558.5 & 8.4(15) & 3909.8 \\
     CH$^+$ & 3.0(09) &  495.9 & 1.5(11) & 2.5(4) & 3.3(12) & 5.4(5) & 2.4(13) & 4.0(6) \\
    CH$_2^+$ & 2.4(09) &   15.9 & 1.3(11) &  826.8 & 2.4(12) & 1.6(4) & 6.5(12) & 4.2(4) \\
     OH$^+$ & 2.2(09) &    1.1 & 1.5(10) &    7.3 & 1.3(12) & 651.9 & 1.8(13) &   8931.1 \\
  H$_3$O$^+$ & 9.0(13) &    1.3 & 8.1(13) &    1.2 & 7.0(13) & 1.1 & 6.2(13) &    0.9 \\
     CO$^+$ & 3.1(08) &    1.5 & 4.9(10) &  243.1 & 5.5(11) & 2728.0 & 4.5(11) & 2244.5 \\
   HCNH$^+$ & 3.2(13) &   30.5 & 2.3(13) &   22.2 & 2.1(12) & 2.0 & 6.1(11) &    0.6 \\
    HCO$^+$ & 4.1(14) &    1.1 & 4.3(14) &    1.2 & 5.7(14) & 1.5 & 3.9(14) &    1.0 \\
    HOC$^+$ & 1.1(10) &   27.0 & 3.4(10) &   87.7 & 1.6(12) & 4168.1 & 4.5(11) & 1138.3 \\
     NO$^+$ & 2.6(12) &    2.8 & 3.2(12) &    3.5 & 1.9(12) & 2.1 & 1.0(12) &    1.1 \\
    HCS$^+$ & 1.9(12) &    3.8 & 1.8(12) &    3.5 & 9.5(11) & 1.9 & 7.8(11) &    1.5 \\ 
     SO$^+$ & 6.5(12) &   24.3 & 1.3(13) &   47.1 & 5.8(12) & 20.8 & 1.4(12) &    5.2 \\ \hline
         CO & 3.9(19) &    1.0 & 3.9(19) &    1.0 & 3.8(19) & 1.0 & 3.7(19) &    1.0 \\
     H$_2$O & 4.5(18) &    1.1 & 4.2(18) &    1.0 & 3.6(18) & 0.8 & 3.0(18) &    0.7 \\
     CO$_2$ & 1.2(18) &    0.9 & 1.2(18) &    0.9 & 1.1(18) & 0.9 & 1.1(18) &    0.9 \\
      N$_2$ & 7.3(18) &    1.0 & 7.3(18) &    1.0 & 7.3(18) & 1.0 & 7.2(18) &    1.0 \\ \hline 
\end{tabular}
\begin{list}{}{}
\item[] a(b) means $a \times 10^b$. The ratio is the enhancement relative to the model without an inner UV source, i.e. $N(G_{0,\rm{in}} = 10^x)/N(G_{0,\rm{in}} = 0)$
\end{list}
\end{table*}

\begin{figure*}
\centering
\includegraphics[width=17cm]{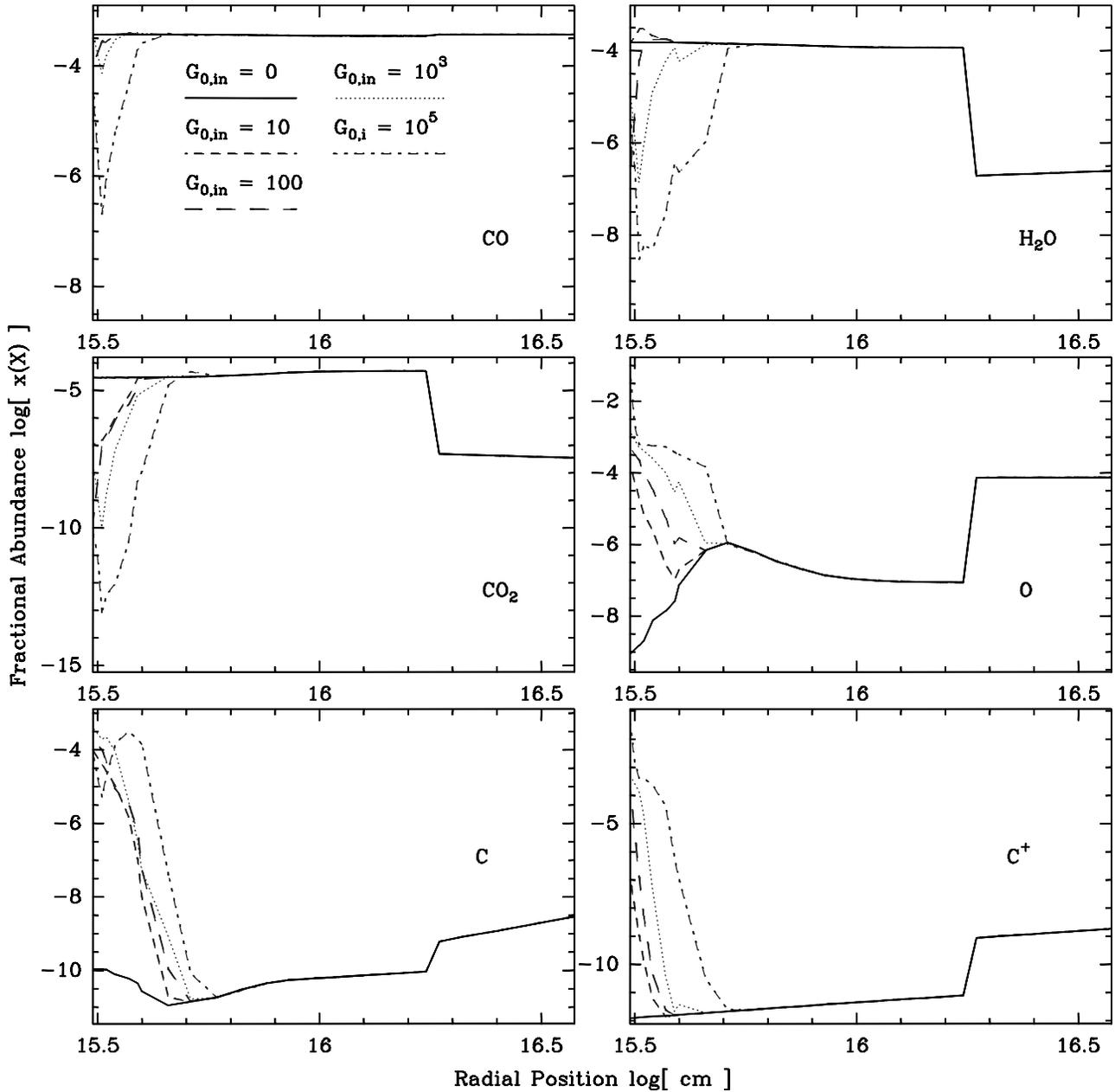}
\caption{Depth-dependent fractional abundances of the basic molecules \element{CO}, \element{H_{2}O} and \element{CO_2} and the `PDR-related' species \ion{O}, \element{C}  and \element{C^+} for different UV fluxes.}
\label{fmajor}
\end{figure*}

\begin{figure*}
\centering
\includegraphics[width=17cm]{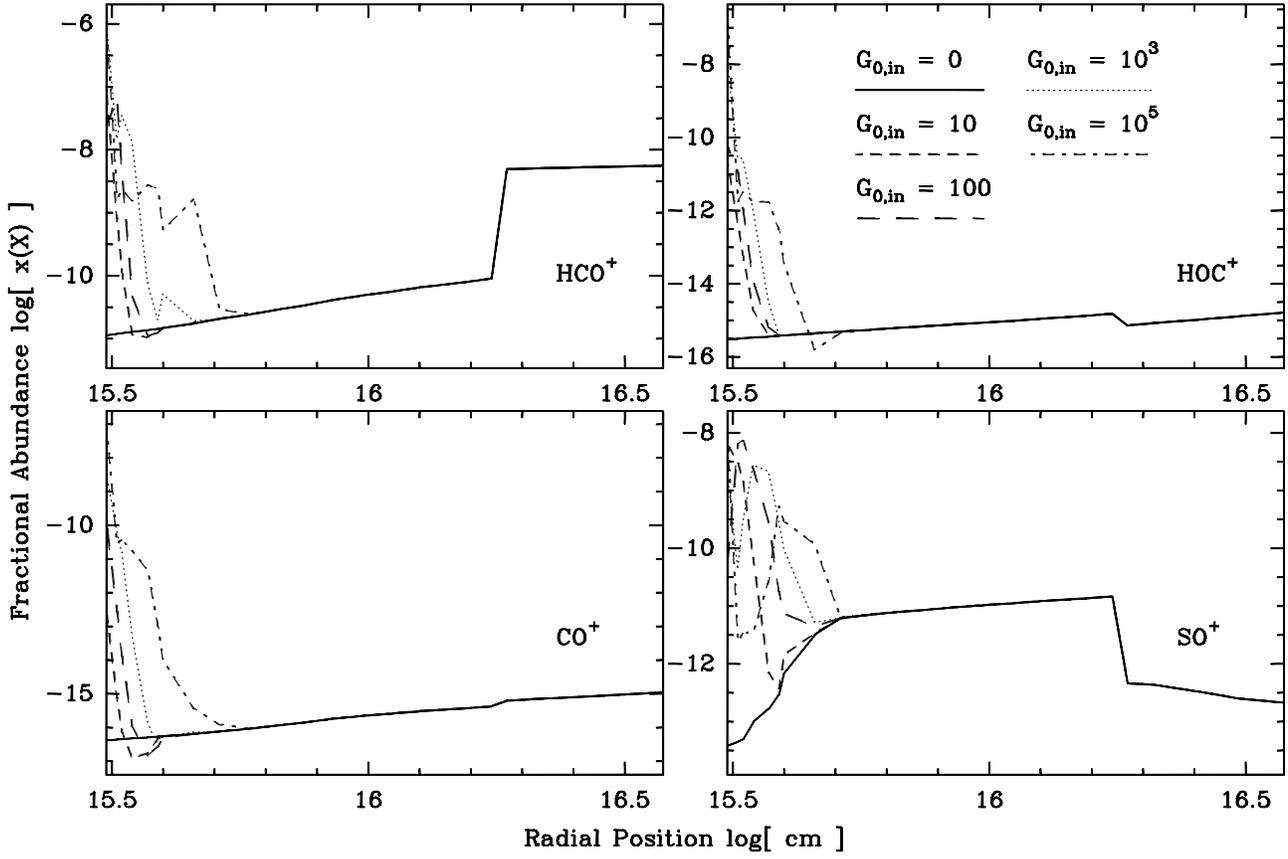}
\caption{Depth-dependent fractional abundances of \element{HCO^+}, \element{HOC^+}, \element{CO^+} and \element{SO^+} for different UV fluxes.}
\label{frion}
\end{figure*}

\begin{figure*}
\centering
\includegraphics[width=17cm]{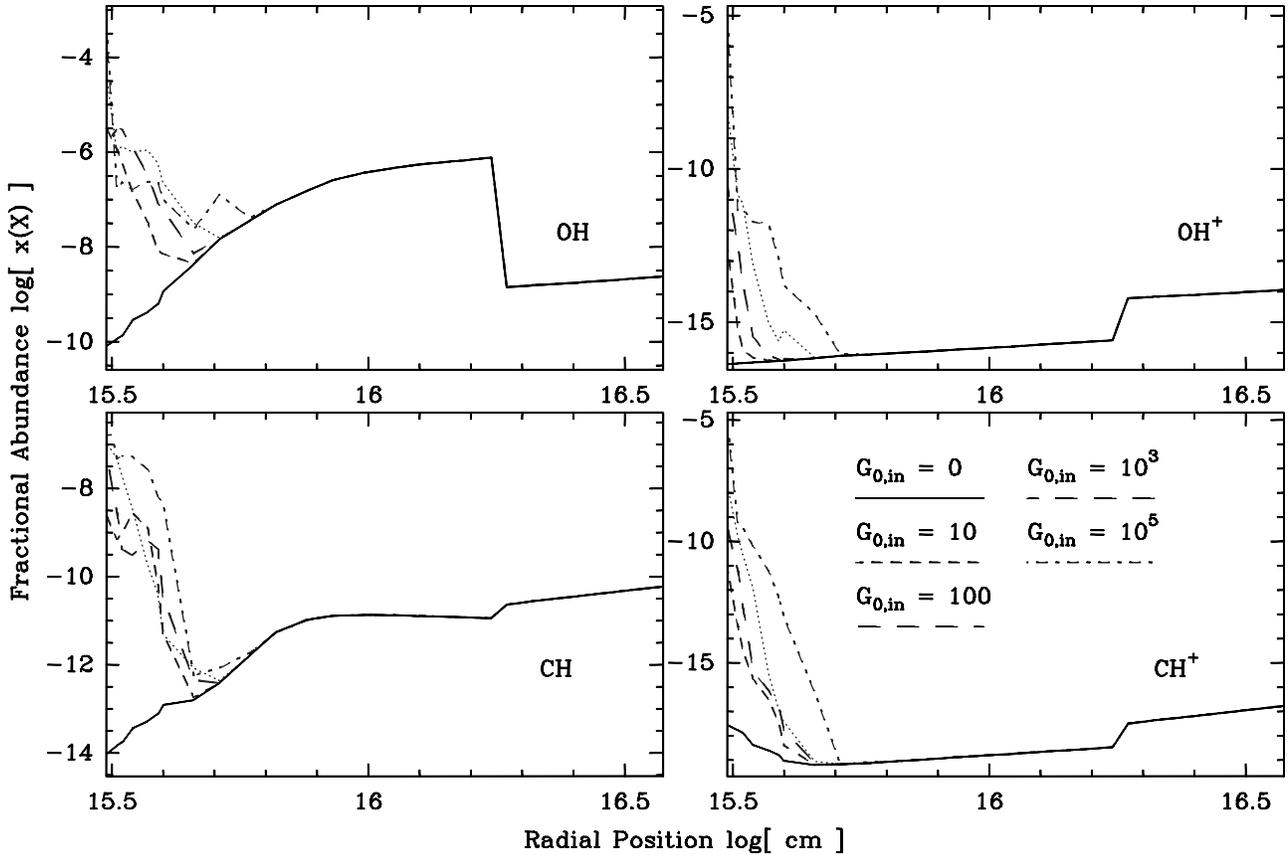}
\caption{Depth-dependent fractional abundances of the hydrides \element{OH}, \element{OH^+}, \element{CH} and \element{CH^{+}} for different UV fluxes.}
\label{fhifi}
\end{figure*}

\begin{figure*}
\centering
\includegraphics[width=17cm]{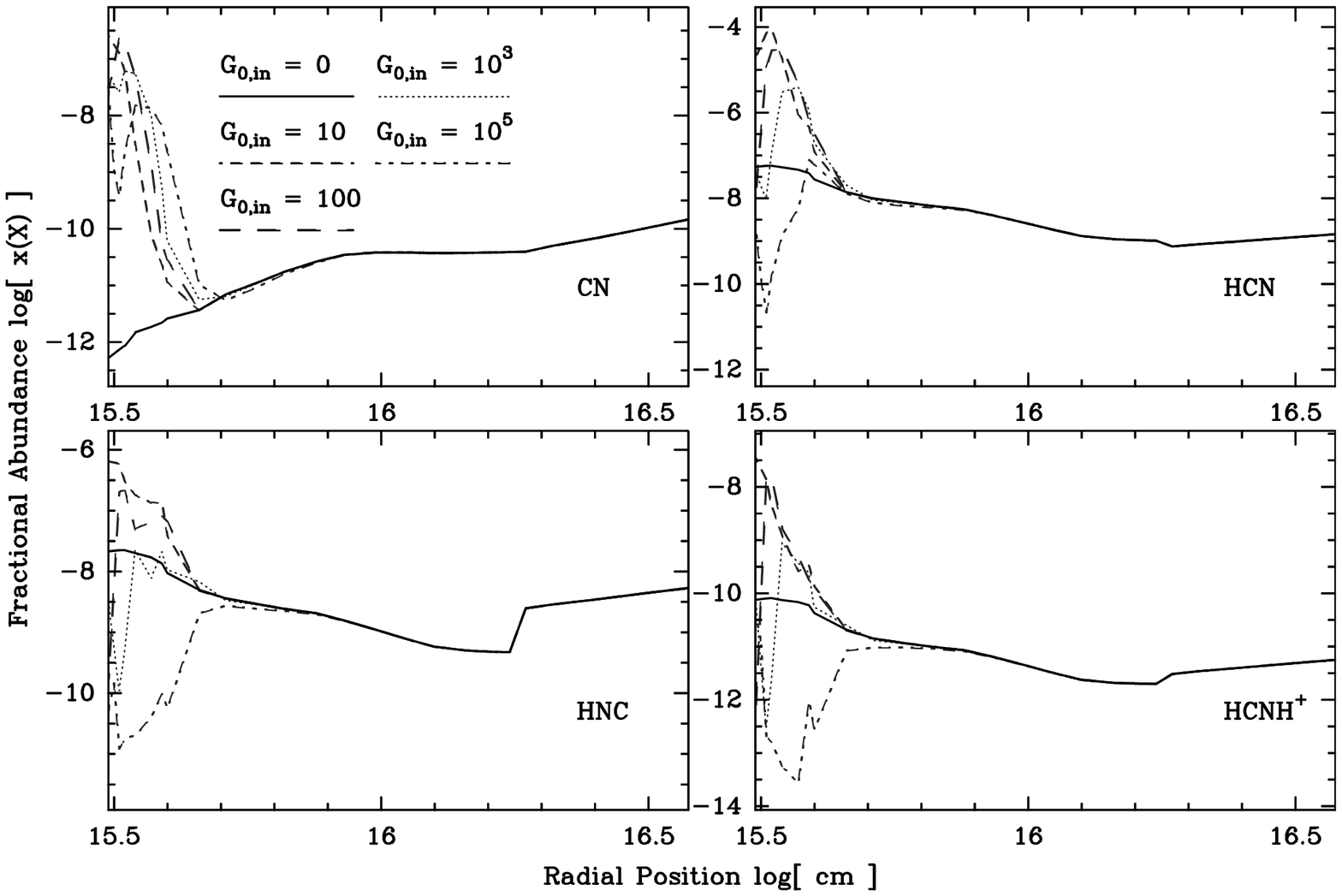}
\caption{Depth-dependent fractional abundances of \element{CN}, \element{HCN}, \element{HNC} and
\element{HCNH^+} for different UV fluxes.}
\label{fcn}
\end{figure*}

We have modeled the envelope around AFGL $2591$ with different UV
fluxes $G_{0,\rm{in}}$ at the inner boundary and compared our results
to those of Doty et al. (\cite{Doty}), representing the model
$G_{0,\rm{in}} = 0$ in this paper. Photodissociation and
photoionization processes are generally fast for the main species in
the regions where they are significant ($\approx 10^{-8}$\,s$^{-1}$)
so that the effects are not significantly different for
$3 \times 10^3$ and $3 \times 10^5$ years. The results are therefore
presented for $t = 3 \times 10^{4}$\,years, the chemical age of AFGL
$2591$ as proposed by Doty et al. (\cite{Doty}).  Total column
densities for a selection of species are predicted in Table
\ref{tcd}. The selection criteria are: (i) the species have a column
density larger than $N(\rm{X}) = 10^{10}$\,cm$^{-2}$; (ii) the
relative difference of the total column densities with and without an
inner UV source is more than $\approx 30$\%; and (iii) the species are
observable with present or future instrumentation. In addition we show
the abundances of the model predictions for basic molecules (other
than \element{H_2}) such as \element{CO}, \element{H_2O}, \element{CO_2} and
\element{N_2} at the bottom of Table~\ref{tcd}.  In general, the
species whose abundances are particularly enhanced are the same as
those found in traditional PDR models, i.e., radicals and ions, and
their chemically related species. The reason that most ratios given in Table~\ref{tcd} 
are $> 1$ is that small differences in very abundant species like CO or H$_2$O can 
make large differences in much less abundant species.

In Fig.~\ref{fdens}, the column densities for various molecules as
functions of $G_{0,\rm{in}}$ are presented, normalized to the values
for $G_{0,\rm{in}} = 0$. Figs.~\ref{fmajor} to \ref{fcn} show the
depth-dependent fractional abundances for selected species in the
inner $2500$\,AU. The radial scale is taken to be somewhat smaller in
these plots than the full extent of the envelope ($29\,000$\,AU,
$\log(r) = 17.6$) in order to emphasize the region which is influenced most
by the inner UV field (see Doty et al. \cite{Doty} for variations in
the outermost regions). Even for the highest values of $G_{0,{\rm
in}}$, the effects of the UV radiation do not penetrate beyond the
freeze-out radius at $100$\,K ($\log(r) = 16.2$).

 Several molecules shown in Figs.~\ref{fmajor} to \ref{fcn} have local minima or maxima 
at certain positions (for example H$_2$O at $\log(r) \approx 15.5$ for $G_{0,\rm{in}} = 10$ 
or HCO$^+$ at $\log(r) \approx 15.7$ for $G_{0,\rm{in}} = 10^5$). These peaks 
are not caused by numerical artifacts but are due to different chemical affects, some of
which are explained in the following sections.

Table~\ref{tcd} lists the radial column densities $N_{\rm{radial}} =
\int n(r)\,dr$. For comparison of observed emission lines to
model-predicted column densities, beam-averaged column densities
defined by $N_{\rm{beam}} \equiv \iint n(z,p)\,G(p)\,2\pi\,p\,dp\,dz$
/ $\int G(p)\,2\pi\,p\,dp$ have to be calculated, where $p$ is the
impact parameter and $G(p)$ is the beam response
function.  Beam averages are centered on the source whereas the calculated radial column 
densities are through the cloud center. $N_{\rm{beam}}$, generally has slightly lower 
values than $N_{\rm{radial}}$ for typical observing beams. Rather than calculating
$N_{\rm beam}$, however, we have computed the actual line fluxes by
solving the equations of molecular excitation and radiative transfer
throughout the envelope and convolving the emerging emission with the
telescope beam (see Sect.~\ref{pline}).

\subsection{Basic molecules: \element{CO}, \element{H_2O}, and \element{CO_2}}
\label{mjm}

As seen in Table~\ref{tcd} and Fig.~\ref{fmajor}, the total radial
column densities of the high abundance molecules \element{CO},
\element{H_2O} and \element{CO_2} are not affected significantly by an
inner UV field. The major destruction mechanism for these molecules is
photodissociation. Measurable changes can only be seen for relatively
high UV fields in the dense region within $\approx 500$\,AU from the
central source. For \element{H_2O} another efficient destructive
reaction is the production of \element{HCO^+} and
\element{HOC^+} through reactions with \element{C^+}. The slight
increase in the relative abundance of \element{H_2O} at $r \approx 4 \times 10^{15}$\,cm
for $G_{0,\rm{in}} = 10$ and $G_{0,\rm{in}} = 100$ is caused by
the protonation of \element{HCN} through \element{H_3O^+} which leads
to \element{HCNH^+} and \element{H_2O}, and by the high-temperature
reaction of molecular hydrogen with \element{OH}, also producing
\element{H_2O}. All reactants are slightly increased by a moderate UV
field, enough to speed up these reactions. At $\approx 5 \times
10^{15}$\,cm the relative abundance of \element{CO} is the same for
all UV fields. This is slightly closer to the star than the
enhancement of \element{OH} for high UV fields (see also
Fig.~\ref{fhifi}) and since \element{CO} $+$ \element{OH}
$\rightarrow$ \element{CO_2} $+$ \element{H} is the dominant reaction
to produce gas-phase \element{CO_2}, the enhancement of \element{OH}
leads to the small increase of \element{CO_2} between $5$--$6 \times
10^{15}$\,cm for $G_{0,\rm{in}} \ga 10^{3}$. The sudden decrease of
the \element{H_2O} and \element{CO_2} abundances at $\approx 2 \times
10^{16}$\,cm ($T \approx 100$\,K) is due to the effects of their
freeze-out onto grains (see Doty et al. \cite{Doty} for more
information).  The abundance minima of H$_2$O, CO and CO$_2$ for 
$G_{0,\rm{in}} \ge 10^{3}$ are due to the higher gas temperature for 
these high FUV fields: the abundances are rising again toward the protostar.

\element{CO_2} is a highly abundant molecule in our model with a
column density of the order of $10^{18}$\,cm$^{-2}$. Boonman et
al. (\cite{BoonmanCO2}) derive a total column density of $2.6 \times
10^{16}$\,cm$^{-2}$ for AFGL $2591$ from observations with the {\it Infrared Space
Observatory}. Even high UV fields do not reduce much 
\element{CO_2}. This overprediction of \element{CO_2} has also been
discussed by Doty et al. (\cite{Doty}), and our new models add little
to this discussion: enhanced UV photodissociation does not solve the
problem.

\subsection{PDR-related species: \element{O}, \element{C}, and \element{C^+}}
\label{ppdr}

Fig.~\ref{fmajor} also shows the dominant line emitters, and thus
coolants, in a PDR: \element{O}, \element{C} and \element{C^+}.
Photodissociation of \element{CO} is the main source for both carbon
and oxygen. The major destruction mechanism for carbon in a low
($G_{0,\rm{in}} = 10$) UV field is the reaction with \element{OH}
leading to \element{CO} and atomic hydrogen. Photoionization of carbon
becomes important for higher UV fields and is the main production
channel of \element{C^+}. The somewhat unexpected jump of
\element{C^+} at $\approx 2 \times 10^{16}$\,cm can be explained by the
freeze-out of \element{H_2O}: the main destroyer of \element{C^+} at
this distance is depleted onto grains below $100$\,K. The relatively
high atomic oxygen abundance for $T < 100$\,K is due to our initial
elemental abundance, consistent with Meyer et al. (\cite{Meyer}) for
diffuse clouds.

 The main production channel for \element{O^+} is not the
photoionization of atomic oxygen (which cannot occur for photon
energies less than $13.6$\,eV) but the reactions \element{H^+} $+$
\element{O} $\rightarrow$ \element{O^+} $+$ \element{H} and \element{C^+} $+$
\element{O_2} $\rightarrow$ \element{O^+} $+$ \element{CO}.
 
The observed column density of atomic carbon toward AFGL $2591$ ($N \le
6.8 \times 10^{17}$\,cm$^{-2}$, see Doty et al. \cite{Doty}) can be
reproduced with a UV field of $G_{0,\rm{in}} \ge 10$. The high
\element{C^+} abundances ($N \le 6.8 \times 10^{17}$\,cm$^{-2}$) are
consistent with a UV field of $G_{0,\rm{in}} \ge 10^3$.  However, as
discussed in Sect.~\ref{pline}, the observed antenna temperatures
require high abundances of C and C$^+$ to be extended over a larger
region than just the inner few hundred AU. Thus, these species likely
trace the outer radiation field, $G_{0,\rm{out}} \approx 10-100$,
rather than an inner UV field.

\subsection{Reactive ions}
\label{prion}

From Table~\ref{tcd}, it is seen that \element{HOC^+} is much more
affected by the strength of the UV field than \element{HCO^+}. The
reason for this is that in our models \element{HOC^+} is basically
formed by the two reactions \element{H_3^+} $+$ \element{CO}
$\rightarrow$ \element{HOC^+} $+$ \element{H_2} and \element{C^+} $+$
\element{H_2O} $\rightarrow$ \element{HOC^+} $+$ \element{H}, which
are both highly dependent on the ionization fraction and thus on the
incoming UV field. In contrast more production channels are possible
for \element{HCO^+}, e.g., \element{HOC^+} $+$ \element{H_2}
$\rightarrow$ \element{HCO^+} $+$ \element{H_2}, thus more
\element{HCO^+} than \element{HOC^+} is produced. Another reason for
the much higher abundances of \element{HCO^+} is the jump at $T =
100$\,K where H$_2$O freezes out: since \element{HCO^+} $+$
\element{H_2O} $\rightarrow$ \element{H_3O^+} $+$ \element{CO} is the
main destruction channel, much more \element{HCO^+} remains in the
region below $100$\,K. Water is important, however, to produce
\element{HOC^+}, hence the HOC$^+$ abundance is lower once water has
depleted onto grains. 

 The \element{HCO^+}/\element{HOC^+} abundance ratios
predicted by our chemical models are in the range of $\approx 10^2$--$10^6$ which is at 
the upper limit of the ratios observed toward PDRs.
Fuente et al. (\cite{Fuente}) reported ratios of $50$--$120$ toward 
the reflection nebula NGC $7023$ --- the lowest ratio measured so far --- 
up to $450$ toward the planetary nebula NGC $7027$. Our models only 
reproduce such low values for $G_{0,\rm{in}} \ge 10^3$. However, 
the observed peak values of HOC$^+$ toward several PDRs are in the range 
of the predicted column densities for $G_{0,\rm{in}} = 10$--$100$.

Van der Tak et al. (\cite{Tak1999}) find HCO$^+$ abundances of the order of 
$\approx 10^{-8}$, which are well fitted by our models with $G_{0,\rm{in}} = 10$. 
Unfortunately, there are no observations of \element{HOC^+} toward AFGL $2591$ to date. 
The question therefore is whether these low ratios also hold in massive star-forming 
regions.

For low UV fields the reactive ion \element{CO^+} is produced mainly
by \element{He^+} $+$ \element{CO_2} $\rightarrow$ \element{CO^+} $+$
\element{O} $+$ \element{He} and \element{C^+} $+$ \element{CO_2}
$\rightarrow$ \element{CO^+} $+$ \element{CO}. For higher UV fields
the reaction \element{C^+} $+$ \element{OH} $\rightarrow$
\element{CO^+} $+$ \element{H} is the most efficient to build up
\element{CO^+}. Its destruction for all UV fields is mainly by
reactions with \element{H_2}. The most efficient pathway to \element{SO^+} is
\element{S^+} $+$ \element{OH} $\rightarrow$ \element{SO^+}$+$ \element{H}.
Its destruction is caused mainly by dissociative recombinations with electrons.

\subsection{Hydrides}
\label{phifi}

\begin{figure}
\centering
\resizebox{\hsize}{!}{\includegraphics{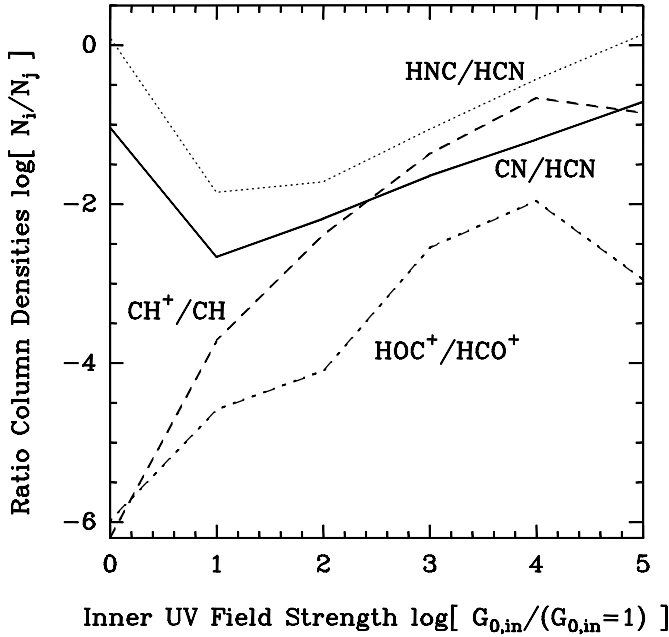}}
\caption{\element{CN}/\element{HCN}, \element{HNC}/\element{HCN}, 
\element{CH^+}/\element{CH} and \element{HOC^+}/\element{HCO^+} column density 
ratios as a function of $G_{0,\rm{in}}$.}
\label{fratio}
\end{figure}

As in traditional PDR models, the abundances of unsaturated hydrides
are strongly increased by UV radiation. \element{OH} and CH$_3$ are
the species that are the least affected by the inner UV field.
\element{CH^+} on the other hand is enhanced by more than a factor of
$10^4$ even for a moderate UV field and its column density grows
steadily with increasing UV field strength. To date, there are
observations only of \element{OH} for AFGL $2591$ reporting $N \ge 4.7
\times 10^{14}$ (see Doty et al. \cite{Doty}). This column density is
well fitted by our models for the region $T < 100$\,K. Our models,
however, produce much more OH in the warmer, inner region (see
Fig.~\ref{fhifi}).  Without UV radiation, \element{OH} is mainly
produced through collisions of atomic oxygen with molecular hydrogen
in the inner warm part. In the presence of a UV field,
photodissociation of H$_2$O becomes the dominant production route. The
OH destruction is primarily through \element{C^+}, and for very high
UV fields ($G_{0,\rm{in}} = 10^{5}$) the photodissociation of
\element{OH} becomes important too. For UV fields with $G_{0,\rm{in}}
\geq 10^{3}$, \element{CH^+} is efficiently produced in the warm inner
regions by reactions of \element{C^+} with molecular hydrogen. For
smaller UV fields, \element{CH^+} is produced mainly by the reaction
\element{HCO^+} $+$ \element{C} $\rightarrow$ \element{CH^+} $+$
\element{CO}. For higher UV fields, photoionization of \element{CH}
becomes the main mechanism for \element{CH^+} production. In all our
models, \element{CH^+} is mainly destroyed in collisions with H$_2$.

For smaller UV fields, \element{CH} is mainly destroyed
through \element{CH} $+$ \element{H_2} $\rightarrow$ \element{CH_2}
$+$ \element{H} and built up by the reverse reaction. \element{OH^+}
is primarily produced by collisions of singly ionized oxygen with
molecular hydrogen. The jump at $\approx 2 \times 10^{16}$\,cm is due
to the increase of atomic oxygen. Charge exchange reactions then
enlarge the \element{OH^+} abundances, in particular the reaction
\element{H_3^+} $+$ \element{O} $\rightarrow$ \element{OH^+} $+$
\element{H_2}. \element{OH^+} is most efficiently reduced in
collisions with \element{H_2}. \element{OH} and \element{OH^+} play a
crucial role in the gas-phase chemistry of water.

Several species such as \element{C^+} and \element{CH^+} are not only
excellent tracers of enhanced UV fields but also of the strength
$G_{0}$ since their abundances rise steadily with increasing field
strengths. The fact that C$^+$ can also be enhanced by an outer UV
field reduces its value as a diagnostic, however. An alternative
tracer is the \element{CH^+}/\element{CH} ratio, which is
$10^3$--$10^5$ times higher if the inner UV is included. The
column density ratio is shown in Fig.~\ref{fratio} as a function of
$G_{0,\rm{in}}$.  Similar arguments hold for other hydrides such as
CH$_2$ and CH$_2^+$ which are less easily observable. Hydrides like CH
and \element{CH^+} are also important for building complex
hydrocarbons (see e.g., van Dishoeck \& Hogerheijde \cite{Vhog});
indeed, a rich carbon chemistry has been observed in PDR layers (e.g.,
Hogerheijde et al. \cite{Hogerheijde95}, Teyssier et
al. \cite{Teyssier}).

It is somewhat surprising that CH$_3$ is much less affected by the
inner UV field than, e.g., CH$_2$. This is only partly true since the
production and destruction mechanisms of CH$_3$ do change by the
impact of the inner UV field. Without the inner UV source CH$_3$ is
efficiently built up by the reaction He$^+$ $+$ CH$_3$OH $\rightarrow$
OH$^+$ $+$ CH$_3$ $+$ He. With the onset of an inner UV field, CH$_3$
is mainly the photodissociation product of species like CH$_3$CN,
CH$_4$ and CH$_3$OH.  CH$_3$ forms H$_2$CO when it reacts with O,
which is the main destruction channel.  Since an inner UV field
clearly enhances atomic oxygen (see Fig.~\ref{fmajor}), this reaction
is very efficient deeper in the cloud where CH$_3$ is destroyed fast.
Thus, although CH$_3$ is first enhanced in the very inner region, it
becomes less abundant deeper in the cloud. This leads to an average
radial column density more or less equal to that where no inner UV
radiation is considered.

\subsection{\element{CN}-bearing species}
\label{pcn}

\begin{figure*}
\centering
\includegraphics[width=17cm]{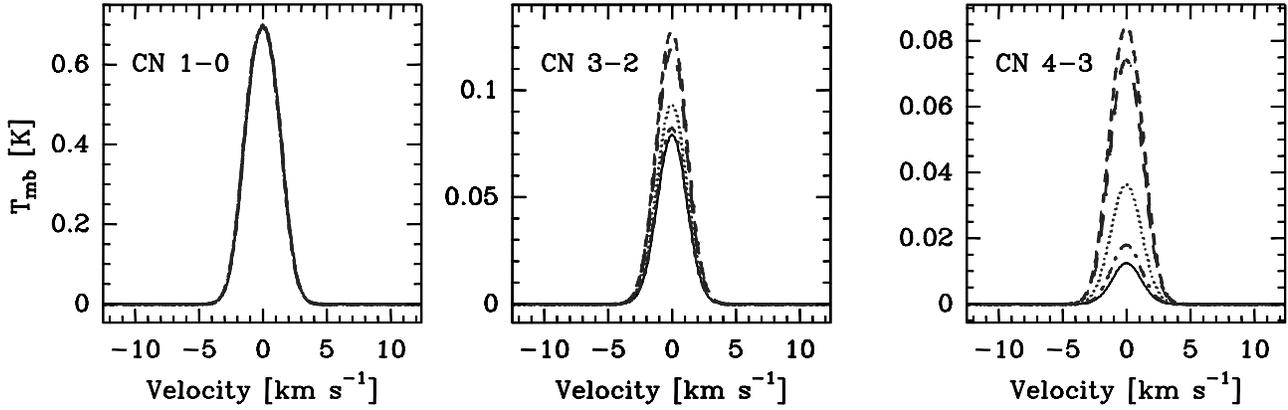}
\caption{CN line profiles calculated for different UV fields. Solid
line: $G_{0,\rm{in}} = 0$. Dashed line (small): $G_{0,\rm{in}} = 10$.
Dashed line (big): $G_{0,\rm{in}} = 100$. Dotted line: $G_{0,\rm{in}}
= 10^{3}$.  Dash-dotted line: $G_{0,\rm{in}} = 10^{5}$. The CN
$1$--$0$ transition has a frequency of $113,491$\,MHz (convolved with
a $33$\arcsec Onsala beam), the \element{CN} $3$--$2$ transition has a
frequency of $340,248$\,MHz (convolved to a $14$\arcsec JCMT beam) and
the \element{CN} $4$--$3$ transition has a frequency of $453,606$\,MHz
(convolved to a $11$\arcsec JCMT beam).  The size of the source is approximately
$30$\arcsec ($29\,000$\,AU).}
\label{fcnline}
\end{figure*}

\begin{figure*}
\centering
\includegraphics[width=17cm]{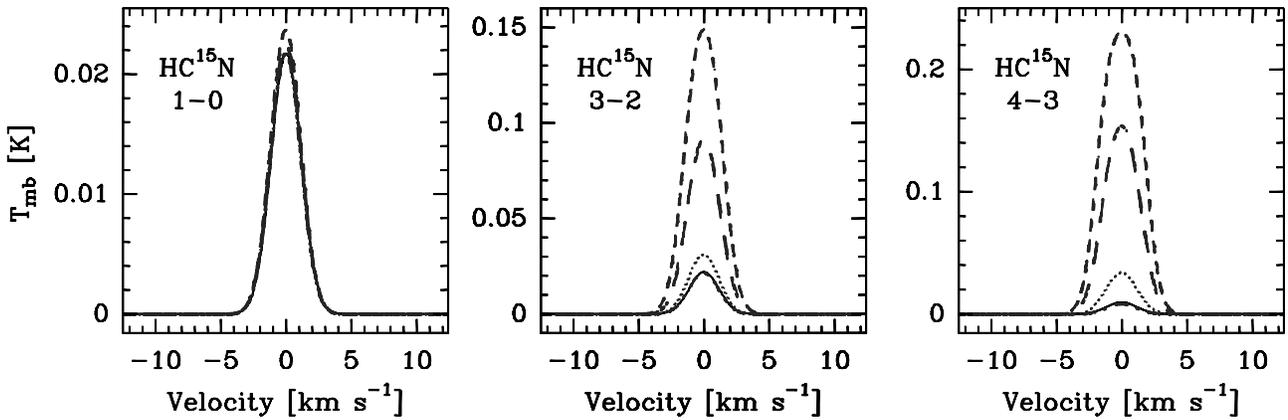}
\caption{HC$^{15}$N line profiles calculated for different UV
fields. Solid line: $G_{0,\rm{in}} = 0$. Dashed line (small):
$G_{0,\rm{in}} = 10$. Dashed line (big): $G_{0,\rm{in}} = 100$. Dotted
line: $G_{0,\rm{in}} = 10^{3}$. Dash-dotted line: $G_{0,\rm{in}} =
10^{5}$. The HC$^{15}$N $1$--$0$ transition has a frequency of
$86$\,GHz (convolved to a $63$\arcsec NRAO beam), the $3$--$2$
transition has a frequency of $258$\,GHz (convolved to a $20$\arcsec
JCMT beam) and the $4$--$3$ transition has a frequency of $344$\,GHz
(convolved to a $14$\arcsec JCMT beam).  The size of the source is approximately
$30$\arcsec ($29\,000$\,AU).}
\label{fhcnline}
\end{figure*}

The chemistry of \element{CN} and \element{HCN} is thought to be
strongly affected by UV radiation and the \element{CN}/\element{HCN}
ratio has been proposed to be a good tracer of PDR chemistry (Fuente
et al. \cite{FuenteCN}, Jansen et al. \cite{Jansen}).
Since our envelopes are more extended than normal
PDRs, we predict lower CN/HCN ratios than those observed in PDRs where
the ratio is typically $\approx 1$ in the shielded regions and $> 1$
in regions exposed to UV radiation. Observations of \element{CN} and
\element{HCN} toward AFGL $2591$ confirm these low ratios. 
The observed abundance of \element{CN} (see Doty et al. \cite{Doty})
is well fitted with $G_{0,\rm{in}} = 10$--$100$.  The \element{HCN} chemistry
is discussed in detail by Doty et al. (\cite{Doty}), especially the
situation in the warm gas-phase ($T > 230$\,K) region where a jump has
been suggested from $x(\rm{HCN}) = 10^{-8}$ to $x(\rm{HCN}) = 10^{-6}$
by Boonman et al. (\cite{BoonmanHCN}).  The observed HCN column
density is reproduced in our models for $G_{0,\rm{in}} = 10$ within a
factor of $2$, whereas it is underproduced in models without an inner UV
field. Alternatively, the model could be extended inward to the radius
where high temperature chemistry may enhance the \element{HCN}
abundances. However, a closer position to the star also means an
increased UV field, and \element{HCN} is more vulnerable to
photodissociation. UV radiation may therefore be an explanation for
the relatively high observed \element{HCN} abundances. The
\element{CN}/\element{HCN} ratio is only enhanced for $G_{0,\rm{in}}
\ga 10^{3}$. For lower UV fields, the ratio is even smaller than that
without UV irradiation (see also Fig.~\ref{fratio}). Hence, the
\element{CN}/\element{HCN} ratio is not a good UV tracer for this high
mass star-forming region, in contrast with the situation for PDRs.

Fig.~\ref{fcn} shows the radial dependences of the abundances of
\element{CN}, \element{HCN}, \element{HNC}
and the reactive molecular ion \element{HCNH^+}. In the warm high density
regime, the production of \element{HCN} through collisions of
\element{CN} with H$_2$ is slightly more efficient than
photodissociation of \element{HCN}, hence more \element{HCN} is
produced than \element{CN}. Collisions of \element{HCN} with
\element{H_3O^+}, however, produce water and \element{HCNH^+}, which
can recombine to form \element{HNC} or \element{CN}. Since this is the main
path to \element{HNC} in our models, more \element{HCN} is produced
compared to \element{HNC}. \element{CH_3CN} is generated through
\element{H_4C_2N^+}, whereas the association of \element{CN} with
\element{CH_3} plays a minor role. \element{CH_3CN} and \element{HNC}
are also photodissociated to \element{CN}. No grain surface production
and subsequent evaporation of CH$_3$CN are assumed. The following
network is valid for UV-affected CN chemistry in our models:

\[
  \begin{array}{ccc}
  \rm{CH_3CN} &
    \stackrel{\rm{H_3^+/H_3O^+}}{\rightleftarrows} &
    \rm{H_4C_2N^+} \\
    \big\downarrow\vcenter{%
    \rlap{$\scriptstyle{\rm{h}\nu}$}} & & 
    \big\uparrow\vcenter{%
    \rlap{$\scriptstyle{\rm{CH_3^+}}$}}\\
    \rm{CN} & \stackrel{\rm{H_2}}{\rightleftarrows} &
    \rm{HCN} \\
    \big\uparrow{%
    \rlap{$\scriptstyle{\rm{h}\nu}$}} & &
    \big\uparrow\big\downarrow\vcenter{%
    \rlap{$\scriptstyle{\rm{H_3O^+/HCO^+}}$}}\\
    \rm{HNC} & \stackrel{\rm{H_3O^+/HCO^+}}{\rightleftarrows} &
    \rm{HCNH^+} \\
  \end{array}
\]

The \element{HNC}/\element{HCN} ratio is $\approx 0.01$--$0.03$ for
low UV fields. The ratio is approximately unity for $G_{0,\rm{in}} =
0$ and slightly higher ($1.4$) for high UV fields ($G_{0,\rm{in}} =
10^5$). The variation of the ratio is mainly due to the changing
\element{HCN} abundances. 
The observed HNC/HCN ratio varies strongly from source to source
but is generally $< 1$. For AFGL $2591$, the observed
ratio is \element{HNC}/\element{HCN}$ \approx 0.01$, which
is in good agreement with a low UV field.

\section{Calculated line emission}
\label{pline}

The temperature, density and abundance profiles presented in
Sect.~\ref{result} have been used as input to the excitation and Monte
Carlo radiative transfer code of Hogerheijde \& van der Tak
(\cite{Hogerheijde}) to compute the emerging line intensities.  The
intrinsic (turbulent) line profile is taken to be a Gaussian with a
Doppler parameter ($1/e$ width) of $1.6$\,km\,s$^{-1}$, independent of
radius. The line profiles are convolved to an appropriate telescope
beam size.  The low-lying transitions trace primarily the low density
outer YSO envelope whereas the higher-lying lines are only excited in
the high density inner envelope. Thus, the intensities of the
higher-lying transitions of species like \element{CN} or
\element{HCN}, whose abundances are affected by UV irradiation, are
expected to be enhanced and the effect on the lines ratios should be
observable.

\subsection{CN and HCN emission lines}

Fig.~\ref{fcnline} shows emission lines of \element{CN} for different
inner UV fields. While the lowest transition \element{CN} $1$--$0$
line has the same intensity for all UV fields, the higher transitions
show clear differences for different UV strengths. The \element{CN}
$3$--$2$ line intensity is increased by a factor of $\approx 2$ for
$G_{0,\rm{in}} = 10$ and $G_{0,\rm{in}} = 100$. The enhancement is
even more noticeable in the \element{CN} $4$--$3$ transition. The
lines for $G_{0,\rm{in}} = 10$ and $G_{0,\rm{in}} = 100$,
respectively, are $\approx 3$ times stronger than that for $G_{0,\rm{in}}
= 10^{3}$, which in turn is approximately twice as strong as that for
$G_{0,\rm{in}} = 10^{5}$ and $G_{0,\rm{in}} = 0$.  Since these
enhancements are clearly larger than the typical $30$\% observational
errors, the impact of central UV photons on the inner envelope is
predicted to have measurable consequences for higher CN transitions.
The line ratios for $G_{0,\rm{in}} = 0$ are \element{CN}
$1$--$0$/$3$--$2$/$4$--$3$ $\approx$ $100$/$11$/$2$, whereas for
$G_{0,\rm{in}} = 10$ they are approximately $100$/$18$/$12$ and for
$G_{0,\rm{in}} = 100$ CN $1$--$0$/$3$--$2$/$4$--$3$ $\approx$
$100$/$17$/$11$.

In Fig.~\ref{fhcnline}, it is seen that the higher-lying
\element{HC^{15}N} lines are similarly enhanced for a moderate UV
field with $G_{0,\rm{in}} = 10$. The optically thin \element{HC^{15}N}
isotope is modeled, since the synthetic \element{HCN} lines are
optically thick and therefore not ideal to show the effects of inner
UV enhancement (see also the discussion by van der Tak et
al. \cite{Tak1999} on HCN modeling).
The \element{HCN}/\element{HC^{15}N} ratio is taken to be $450$
(Wilson \& Rood \cite{Wiro}). The $J=3-2$ line is enhanced by a factor
of $7$ and the $J=4-3$ line a factor of $\approx 26$ for moderate UV
fields. For the lowest transition, all line intensities are the same,
except that for $G_{0,\rm{in}} = 10$, which is approximately $9$\%
higher and thus not distinguishable from the other lines if calibration
errors are taken into account. The intensity ratios for $G_{0,\rm{in}}
= 0$ are \element{HC^{15}N} $1$--$0$/$3$--$2$/$4$--$3$ $\approx$
$10$/$10$/$4$, for $G_{0,\rm{in}} = 10$ 
\element{HC^{15}N} $1$--$0$/$3$--$2$/$4$--$3$ $\approx$
$10$/$64$/$103$ and for $G_{0,\rm{in}} = 100$ 
they are approximately $10$/$41$/$70$. 

Line intensities have also been calculated for an outer UV field of
$G_{0,\rm{in}} = 10$--$100$. It is found, however, that the CN and HCN
lines are not very sensitive to the outer UV field due to the much
lower densities in those regions, with enhancements in line
intensities of only $10$--$20$\%. Thus, the \element{CN} and
\element{HCN} lines trace the central UV field rather than the outer
interstellar radiation field. This is in contrast with species such as
C and C$^+$, whose transitions are readily excited at low densities in
the outer envelope (see below). High spatial resolution interferometry
observations of both low- and high-J CN and HCN lines should be able
to directly constrain the extent over which the inner and outer UV field
affect the chemistry.

\subsection{C and C$^+$ emission lines}

In addition to the molecular lines mentioned above, the line profiles
of the fine-structure transitions of \element{C} and C$^+$ have been
calculated. Although the abundances of both species are strongly
enhanced by the inner UV field, the resulting antenna temperatures are much
lower than observed by van der Tak et al. (\cite{Tak1999}). Thus, even
though the column densities including an inner UV field match those
derived from observations, the beam dilution of the inner region is so
large that the resulting fluxes fall short by $1$--$2$ orders of magnitude.
Therefore, models have also been run in which the outer radiation
field is enhanced above the average ISRF. Such outer enhancements
could be due to other nearby young stars. For example, for AFGL $2591$ a
nearby \ion{H}{ii} region ($6000$\,AU projected distance)
has been found (see e.g., van der Tak et al. \cite{Tak1999}, Trinidad
et al. \cite{Trinidad}), whose exciting star could illuminate (part
of) the outer AFGL $2591$ envelope.  An outer UV field with
$G_{0,\rm{out}} = 100$ and an enhanced gas temperature due to
photoelectric heating matches the [C I] $492$\,GHz observations best. 
Since C$^+$ is mainly produced through photoionization of atomic carbon (ionization energy 
$11.26$\,eV), the same issue may be true for the [C II] line.

\section{Summary and conclusions}
\label{pcon}

\begin{figure}
\centering
\resizebox{\hsize}{!}{\includegraphics{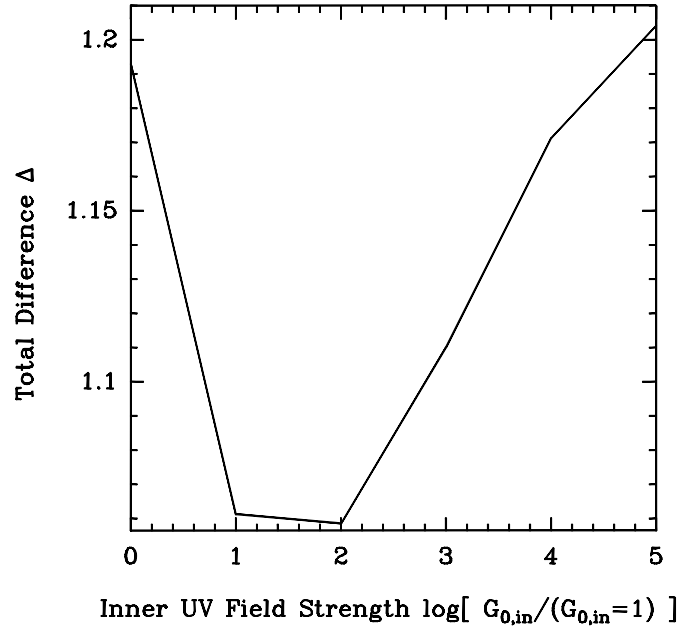}}
\caption{The quality of the model fit to the observations as a function of $G_{0,\rm{in}}$.}
\label{fstat}
\end{figure}

We have extended the detailed chemical models of Doty et
al. (\cite{Doty}) to study the influence of UV irradiation from AFGL
$2591$ on the chemistry in the surrounding envelope on
$200$--$29\,000$\,AU scales. From Figs. \ref{fmajor} to \ref{fcn}, it
is seen that an inner UV flux affects only the region within
$500$--$600$\,AU from the star. The species whose abundances are most
enhanced are radicals and ions, similar to the case of normal PDRs.
However, the chemistry differs from that in normal PDRs in various
details owing to the higher temperature and higher H$_2$O abundance in
the inner YSO envelope.

By comparing our results to those of Doty et al. (\cite{Doty}), we
find that agreement with the observed column densities is improved
with a moderate UV field of $G_{0,\rm{in}} =10$--$100$. This is seen
quantitatively in Fig.~\ref{fstat}, which shows the absolute
differences between the model results and the observations, $\Delta =
\frac{1}{N} \sum_{i=1}^{N}|\log(N_{\mathrm{obs}})-\log(N_{\mathrm{mod}})|$, summed over
all species. Species like CN and HCN are particularly improved with a
modest UV field. However, the high temperature chemistry of
\element{HCN} in the inner region still needs further
investigation. The \element{CN}/\element{HCN} ratio is found not to be
a good tracer for $G_{0,{\rm in}}$ for AFGL $2591$. For low UV fields,
the extra CN reacts with H$_2$ to HCN in the warm, dense region thus
enhancing HCN more than CN, in contrast to typical PDRs.  However,
enhanced intensities of the $3$--$2$ and $4$--$3$ lines of
\element{CN} and \element{HC^{15}N} are predicted compared with the
lower $J=1-0$ lines, which should be observable both with single
dishes and with submillimeter interferometers. Another strong
indicator for enhanced UV fields is the \element{CH^+}/\element{CH}
ratio with values between $10^{-3}$ and $10^{-1}$. Neither species can
readily be observed with ground based observatories, but are excellent
targets for Herschel-HIFI. 

Most other species have their maximum column density at lower UV
fields (see Table~\ref{tcd}). This is due to different preferred
chemical networks for each strength of the UV field; in particular,
many species are photodissociated away for very high UV fields. Like
in Doty et al. (\cite{Doty}), the \element{CO_2} column density is
overpredicted by a factor of $40$ in our model.  In this sense, UV
irradiation from the central star proves not to be an efficient
destruction mechanism for \element{CO_2}. Impulsive heating events and
X-ray chemistry may destroy carbon dioxide more efficiently.

Our model results indicate strongly that UV radiation from the high
mass YSO AFGL $2591$ cannot be neglected in the surrounding envelope
chemistry. The inferred value of $G_{0,{\rm in}} = 10$--$100$ for an
inner UV field is reasonable for a young massive star like AFGL $2591$
and the corresponding column density and dust opacity can be justified
by the mass inflow rate  even though we do not present a fully 
self-consistent model (see Sect.~\ref{puvf}). Our models
assume a spherical symmetry and neglect geometric effects. High
spatial resolution data of species like CN should also be able to
establish the importance of non-spherical effects, such as the escape
of UV photons through the outflow cones.

In recent years it has become clear that YSOs are also
strong X-ray emitters. The observed X-ray luminosities range from
approximately 10$^{28.5}$ to 10$^{33}$\,erg\,s$^{-1}$ (Feigelson \&
Montmerle \cite{Feig}, Preibisch \cite{Preibisch}). X-rays can also
penetrate deeper into clouds than UV photons due to smaller absorption
cross sections at high energies. Most recently Doty et al. (\cite{Doty04}) suggested that
X-rays may be responsible for a higher ionization rate in the low-mass hot core IRAS $16\,293$--$2422$.
It is therefore important to also investigate the effects of X-rays on the chemistry 
in the envelopes of YSOs and such work is in progress (St\"auber et al. 2004, in preparation).

\section{Acknowledgments}

The authors are grateful to Michiel Hogerheijde and Floris van der Tak for the use of 
their Monte Carlo code.  We thank the anonymous referee for the valuable comments. This work was partially supported under grants from The Research 
Corporation (SDD). Astrochemistry in Leiden is supported by the Netherlands Research School 
for Astronomy (NOVA) and by a Spinoza grant from the Netherlands Organization for 
Scientific Research (NWO).

\end{document}